\documentclass[11pt]{article}
\usepackage{UF_FRED_paper_style}

\title{Nigeria's Digital Sovereignty: Analysis of Cybersecurity Legislation, Policies, and Strategies}

\author{Polra Victor Falade\\
    Department of Cyber Security, \\Nigerian Defence Academy, Kaduna, Nigeria\\
    \href{mailto:pvfalade@nda.edu.ng}{\texttt{pvfalade@nda.edu.ng}} 
\and Oluwafemi Osho\\
    School of Computing, \\Clemson University\\
    \href{mailto:oosho@clemson.edu}{\texttt{oosho@clemson.edu}} 
    }
    
\date{\today}

\begin{document}
{\setstretch{.8}
\maketitle
\begin{abstract}

This paper examines Nigeria’s pursuit of digital sovereignty through two core instruments: the Cybercrimes (Prohibition, Prevention, etc.) Act and the National Cybersecurity Policy and Strategy (NCPS). Despite recent reforms, it remains unclear whether these frameworks effectively secure Nigeria’s digital domain and advance its digital sovereignty amid escalating cross-border cyber threats. Using a multi-method, triangulated qualitative design that combines document analysis, secondary analysis of existing studies, expert insights, and direct observation of cybersecurity developments, the paper assesses how these instruments operate in practice. The Cybercrimes Act (2015, amended 2024) and NCPS (2015, revised 2021) have strengthened Nigeria’s commitments to tackling cybercrime, regulating digital activities, and protecting critical infrastructure. Yet persistent gaps remain, including legislative ambiguities, weak enforcement, uneven threat prioritization, limited institutional coordination, and loss of skilled professionals. The paper argues that achieving digital sovereignty will require stronger implementation, sustainable resourcing, workforce retention, and clearer accountability mechanisms to translate policy ambition into tangible and durable security outcomes.\\


\noindent

\end{abstract}
}


\section{Introduction}
The concept of digital sovereignty is multifaceted and continually evolving, with its exact meaning varying across different national contexts \citep{pohle2020digital,couture2019does}. One of the major connotations relates to a nation’s capacity to govern its digital territory, encompassing the protection of its digital infrastructure and the exercise of authority over digital communications that affect its territory and citizens \citep{pohle2020digital,fleming2025digital}. Today, digital sovereignty has emerged as a crucial strategic instrument for national governments to maintain control over their technological future \citep{yalamancheli2025digital}. 

Cybersecurity constitutes a fundamental pillar of digital sovereignty. As Katsikas \citeyearpar{katsikas2025towards} observes, digital sovereignty entails “both protective mechanisms and offensive tools that drive digital innovation” (p. 279), illustrating the integral role of cybersecurity in realizing digital sovereignty. Cyber threats, which may arise from both within and outside a nation’s geographical and cyber borders, directly undermine its ability to exercise authority and safeguard its digital domain. To address this challenge, enhancing the resilience of critical sectors, processes, and data has become a central requirement for sustaining digital sovereignty \citep{moerel2021reflections}. In practice, nation states have pursued this goal through a range of legal and policy instruments, including the European Union’s General Data Protection Regulation (GDPR) \citep{celeste2021digital}, China’s Cybersecurity Law (CSL), Data Security Law (DSL), and Personal Information Protection Law (PIPL) \citep{creemers2022china}, as well as the United States’ Clarifying Lawful Overseas Use of Data (CLOUD) Act \citep{rutherford2019cloud}. Within developing regions, such as Africa, this effort is reflected in frameworks like the African Union Convention on Cyber Security and Personal Data Protection,\footnote{https://au.int/en/treaties/african-union-convention-cyber-security-and-personal-data-protection.} adopted in June 2014. These examples suggest that securing the digital future of any nation fundamentally depends on its ability to build resilience against cyber threats and exercise sovereign authority over its digital space. Because cyber risks transcend geographical borders, states cannot rely solely on traditional security measures; instead, they must establish robust legal, policy, and institutional frameworks to protect their critical sectors, processes, and data.

Scholars have examined how different nations employ cybersecurity frameworks and initiatives to reinforce their digital sovereignty \citep{qin2025regulatory,roberts2021safeguarding,kianpour2025digital}. Building on this literature, the present paper evaluates two central instruments adopted by the Nigerian government to advance its digital sovereignty: the Cybercrimes Act and the National Cybersecurity Policy and Strategy. Several factors motivated the choice of Nigeria as a case study. First, as Africa’s most populous nation and a regional leader, Nigeria hosts one of the continent’s largest and fastest-growing digital economies \citep{worldbank2019nigeria}. With ICT contributing nearly 20\% of national GDP in recent years \citep{ita2025nigeria}, the country drives a substantial share of West Africa’s digital economy and cyber activity \citep{interpol2025africa}. Second, Nigeria faces a growing incidence of cyber threats and consistently ranks among the most frequently targeted states globally \citep{interpol2025africa,adewopo2024comprehensive,osho2023national,osho2024national,osho2025nigeria}. Given its demographic weight and strategic position, Nigeria’s cybersecurity posture could hold far-reaching implications for continental digital governance and the pursuit of digital sovereignty. 

In response to these opportunities and challenges, the Nigerian government has taken notable steps to entrench cybersecurity governance, for example, through the Cybercrimes Act (2015, amended 2024)\footnote{https://rb.gy/1csckw} and the National Cybersecurity Policy and Strategy (2015, updated 2021).\footnote{https://rb.gy/ypt0nz} These frameworks mark significant progress in articulating a national approach to safeguarding the digital domain and advancing digital sovereignty. However, questions remain about their effectiveness in practice and their capacity to address Nigeria’s evolving cybersecurity landscape. It is against this backdrop that our study evaluates these two instruments. 

The rest of the paper is organized as follows: Section \ref{landscape} discusses the current cybersecurity landscape in Nigeria. In Section \ref{act}, we evaluate the country's Cybercrime Act. Section \ref{policy} focuses on the analysis of the national cybersecurity policy and strategy. We provide recommendations for enhancing national digital sovereignty in Section \ref{recommendations}. The study concludes in Section \ref{conclusion}.

\section{Nigeria's Cybersecurity Landscape}
\label{landscape}
Nigeria’s rapid digital transformation has expanded its economic and social frontiers, offering immense opportunities while simultaneously introducing profound cybersecurity challenges with implications for digital sovereignty. Driven by advances in information and communication technology (ICT), Internet adoption in Nigeria has grown exponentially. The number of Internet users surged from 45 million in 2011 to over 115 million by 2021 and is projected to reach 143 million by 2026 \citep{george2022cybersecurity}. This growth, coupled with the proliferation of fintech innovations, e-government initiatives, and the African Continental Free Trade Area (AfCFTA) digital trade prospects, has positioned Nigeria as a potential leader in Africa’s digital economy \citep{abdullahi2021historical}.

Yet, this rapid digital expansion has been paralleled by rising cyber threats that undermine national security, economic stability, and public trust. Cybercrime in Nigeria manifests in various forms, including phishing, ransomware, identity theft, misinformation, and the notorious advance-fee fraud or “419 scams” \citep{abdullahi2021historical,chawki2009nigeria}. The Nigeria Deposit Insurance Corporation (NDIC) reported a 183\% increase in e-payment fraud within the banking sector between 2013 and 2014, while cybercrime losses in 2014 alone amounted to 127 billion Naira, approximately 0.08\% of the national GDP \citep{abdullahi2021historical}. These attacks not only erode confidence in Nigeria’s financial systems but also impose reputational costs, with Nigerian Internet Service Providers (ISPs) frequently blacklisted by global email systems \citep{awhefeada2020appraising}.

The institutional framework for cybersecurity in Nigeria includes the Office of the National Security Adviser (ONSA), which houses the National Computer Emergency Response Team (ngCERT), the Nigerian Communications Commission (NCC), the National Information Technology Development Agency (NITDA), law enforcement agencies, and the judiciary \citep{george2022cybersecurity}. Despite this institutional presence, enforcement bottlenecks persist. Investigations and prosecutions are hampered by inadequate technical capacity, slow judicial processes, and limited inter-agency coordination \citep{abdullahi2021historical}. These weaknesses hinder Nigeria’s ability to assert control over its digital space, thereby constraining its digital sovereignty.

Socioeconomic factors compound the problem. Studies identify poverty, unemployment, corruption, and low digital literacy as root causes of cybercrime in Nigeria \citep{abdullahi2021historical,hassan2012cybercrime}. Secondary drivers include weak policy implementation, negative role models, and the pursuit of wealth and social status \citep{adeniran2008internet}. The prevalence of “yahoo yahoo” scams, conducted through platforms such as Yahoo Mail and social media, exemplifies how social and economic motivations intersect with digital opportunities to fuel cybercrime \citep{chawki2009nigeria}.

Moreover, the transnational nature of cybercrime complicates enforcement. Many perpetrators operate across borders, exploiting gaps in international legal cooperation. Scholars argue that without harmonised cyber laws and mutual legal assistance frameworks, prosecutions remain limited \citep{broadhurst2012cybercrime,olowu2009cybercrimes}. For Nigeria, aligning domestic legislation with global standards is essential not only for tackling cybercrime but also for asserting sovereignty in cyberspace, where jurisdictional boundaries are often blurred \citep{umejiaku2016legal}.



\section{Research Methodology}

This study employed a multi-method, triangulated qualitative design to examine how Nigeria’s key cybersecurity frameworks function in practice and how far they advance the country’s digital sovereignty. Specifically, the research combined qualitative document analysis, secondary analysis of existing studies, expert elicitation, and direct observations of developments in the Nigerian cybersecurity landscape. This combination of methods enabled a comprehensive assessment of persistent challenges and gaps that remain unaddressed despite recent legislative reforms.

The core of the analysis focused on two key cybersecurity legal and policy instruments: (i) The Cybercrimes (Prohibition, Prevention, etc.) Act 2015 and its 2024 amendments, and (ii) The National Cybersecurity Policy and Strategy (NCPS), both its initial and updated versions. We analyzed the documents to identify their scope, enforcement mechanisms, regulatory obligations, institutional arrangements, and areas targeted for reform. In addition, we reviewed official reports, policy briefs, implementation guidelines, and scholarly critiques related to these instruments. These secondary sources were used to assess real-world outcomes, institutional performance, and diverse stakeholder perspectives on the effectiveness and limitations of Nigeria’s cybersecurity regime. The analyses examined the extent to which the Cybercrimes Act and the NCPS provide coherent, enforceable, and context-appropriate mechanisms for addressing evolving cybersecurity threats.

Furthermore, we drew on our expert insights and professional experience within the Nigerian cybersecurity ecosystem to interpret the practical implications of these instruments. These expert perspectives informed the discussion of how the legal and policy frameworks shape, enable, or constrain Nigeria’s capacity to exercise meaningful control over its digital space. 

To strengthen the validity of our findings, we triangulated the documentary and secondary data with direct observations of current cybersecurity practices and institutional responses within Nigeria’s digital ecosystem. These observations included ongoing developments in incident response, sectoral regulations, inter-agency collaboration, and public–private engagement around cybersecurity. By comparing what the instruments prescribe with how institutions and stakeholders actually behave in practice, we were able to further identify misalignments, implementation gaps, and emerging challenges.

\section{Cybercrimes (Prohibition, Prevention, etc) Act}
\label{act}

\subsection{Overview and Key Provisions of the Cybercrimes Act}
The Nigerian Cybercrimes (Prohibition, Prevention, etc.) Act 2015\footnote{https://rb.gy/ocurqk} represents the country’s first comprehensive federal legislation specifically designed to regulate criminal activities within cyberspace \citep{awhefeada2020nigerian,george2022cybersecurity}. By addressing offences, omissions, and threats arising in the digital domain, the Act established a legal foundation for the prohibition, prevention, detection, and prosecution of cybercrimes and related activities across Nigeria. Its enactment marked a significant milestone in the country’s pursuit of digital sovereignty, as it provided the state with the authority to govern its digital infrastructure, protect critical information systems, and assert jurisdiction over cyber-related offences occurring within or affecting its territory.

Structurally, the Act comprises 59 sections organised into eight key parts: Object and Application; Protection of Critical National Information Infrastructure; Offences and Penalties; Duties of Financial Institutions; Administration and Enforcement; Arrest, Search, Seizure, and Prosecution; Jurisdiction and International Cooperation; and Miscellaneous Provisions. Together, these sections articulate a framework through which Nigeria seeks to enhance control over its cyberspace, strengthen institutional capacity, and coordinate responses to cyber threats both domestically and internationally. The categorisation of these sections, presented in Table 1 of this study, reflects the multidimensional nature of digital sovereignty, encompassing legal authority, technical protection, and cooperative governance mechanisms necessary to secure the nation’s digital landscape.

The 2024 amendment to Nigeria’s Cybercrime Act introduced substantive and procedural reforms aimed at strengthening the country’s capacity to govern and secure its digital ecosystem. In total, twelve sections were revised to improve legislative clarity, enhance institutional capacity, expand regulatory scope, and promote better inter-agency coordination.

Notably, sections 17a, 22(1), and 37 were amended to correct semantic and typographical inconsistencies, including the rectification of the term genuineness in section 17a, thereby improving the legal precision of the Act. More significantly, sections 21 and 41 were revised to bolster the functionality of the National Computer Emergency Response Team (ngCERT). Key changes included reducing the cyber incident reporting window from seven days to 72 hours, mandating the routing of internet traffic through sectoral Security Operations Centres (SOCs), and establishing sectoral Computer Emergency Response Teams (sCERTs). These measures are directly tied to digital sovereignty, as they enhance the state’s technical and institutional capacity to detect, respond to, and mitigate cyber threats within its jurisdiction.

The amendment also addressed evolving cybersecurity risks in critical areas. Section 24, which previously focused on cyberstalking, was expanded to criminalise the dissemination of pornographic materials and knowingly false messages intended to incite lawlessness or violence. While this strengthens the state’s regulatory authority over harmful digital content, civil society groups have expressed concern about the potential for arbitrary enforcement and its implications for privacy and free expression \citep{aljazeera2024}. Similarly, Section 30 broadened its coverage from Automated Teller Machines (ATMs) and Point-of-Sale (PoS) devices to include all forms of payment technology, ensuring the law remains responsive to technological innovation in the financial sector.

In addition, Section 38(1) was amended to align data retention obligations with the Nigeria Data Protection Act (NDPA), although the two-year retention period remained unchanged due to the NDPA’s lack of specificity on timelines. Finally, Section 44 raised the cybersecurity levy on electronic transactions from 0.005\% to 0.5\%, significantly increasing funding for the National Cybersecurity Fund. While this provision strengthens Nigeria’s fiscal capacity to invest in cybersecurity infrastructure—an essential pillar of digital sovereignty, it also raises concerns about potential financial burdens on small and medium-sized enterprises \citep{freedomhouse2024emerging}.

\begin{figure}
\centering
    \includegraphics[width =0.75\textwidth]
{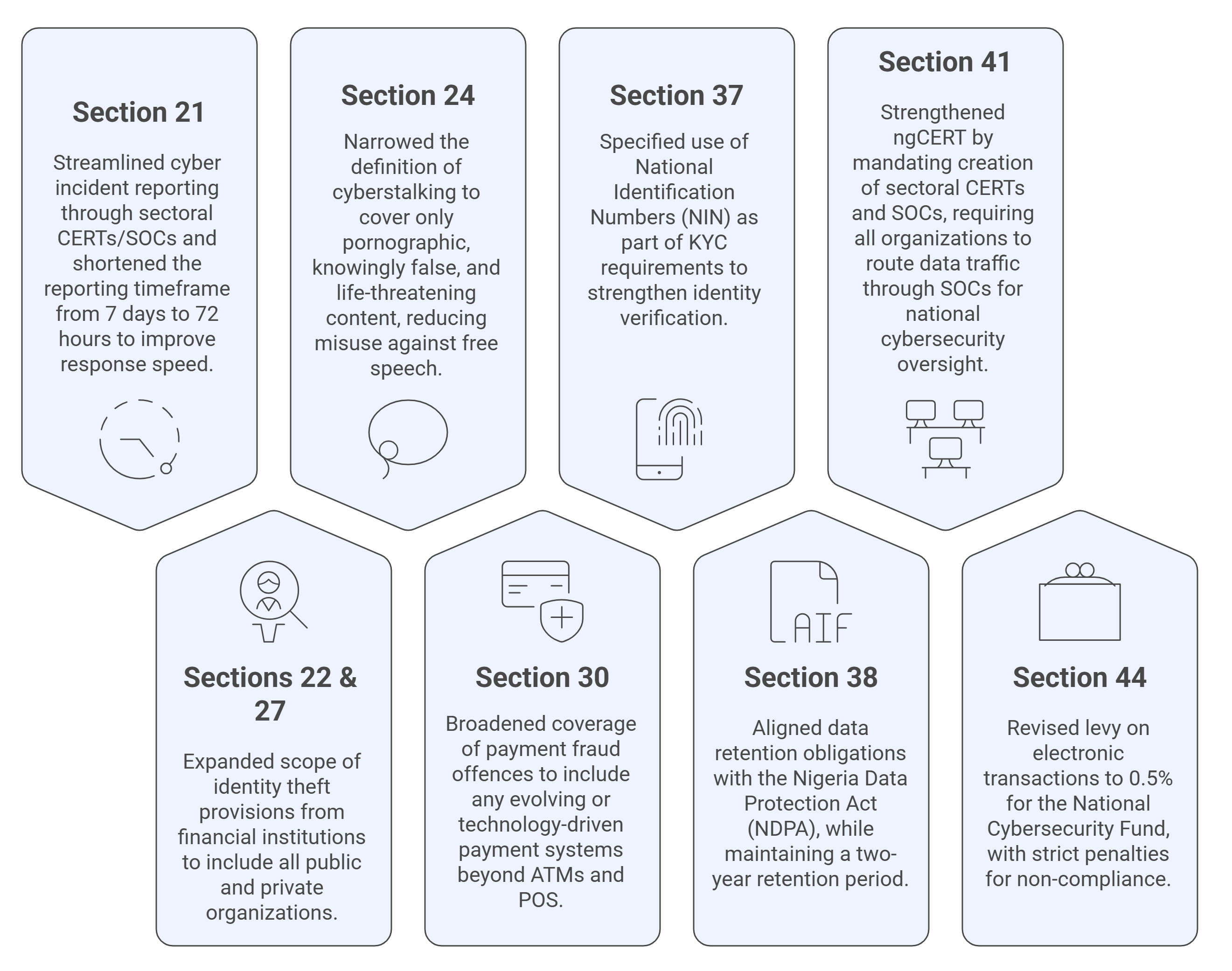}
 \caption{The key amendments in the Cybercrimes (Prohibition, Prevention, etc.) (Amended) Act, 2024.}
 \label{fig:amendment}
 \end{figure}

The amendments represent a deliberate effort to consolidate Nigeria’s control over its digital infrastructure, align regulatory practices with technological advances, and secure the resources and institutions necessary for an autonomous and resilient digital ecosystem. The principal legislative changes are summarized in Figure \ref{fig:amendment}.

\subsection{Impact on Sovereignty}
The enactment of the Cybercrimes (Prohibition, Prevention, etc.) Act 2015 marked a critical milestone in Nigeria’s assertion of digital sovereignty. By introducing a comprehensive legal framework for addressing cybercrime, the Act represented a formal recognition of cyberspace as a domain over which the Nigerian state exercises authority. Before its adoption, cybercrime investigations and prosecutions relied on fragmented legal provisions ill-suited for the digital era. The 2015 Act therefore expanded the state’s regulatory capacity into the digital sphere, empowering institutions to protect critical information infrastructure, prosecute online fraud, combat identity theft, and safeguard citizens from emerging cyber threats. This legal foundation enabled Nigeria to begin consolidating control over its digital ecosystem, a prerequisite for sovereignty in the modern information age.

Furthermore, the Act facilitated Nigeria’s engagement in international cooperation on cybercrime. By criminalising cyber offences in alignment with global standards, the law created a basis for mutual legal assistance treaties and cross-border digital evidence sharing. This was particularly significant because cyber threats frequently transcend national borders, and without such a framework, Nigeria would have lacked the legal mechanisms to request or provide assistance in investigations involving foreign actors. The law also spurred the establishment of specialised cybercrime units, technical infrastructure, and capacity-building initiatives, thereby enhancing the institutional foundations of Nigeria’s sovereignty in cyberspace.

The 2024 amendment to the Act sought to modernise and strengthen these gains. By clarifying certain offences, updating outdated provisions, and introducing measures such as cybersecurity levies to fund enforcement, the amendment provided Nigeria with sharper legal and fiscal tools for digital governance. These changes promised to bolster Nigeria’s ability to protect its digital borders, regulate the activities of global technology companies operating locally, and ensure that resources were available to develop indigenous cybersecurity infrastructure. In this sense, the amendment marked a step toward greater autonomy and resilience in the face of rapidly evolving cyber threats.

The Cybercrime Act and its 2024 amendment represent significant steps towards consolidating Nigeria’s digital sovereignty by providing the legal, institutional, and fiscal mechanisms necessary to govern the digital domain. However, the extent to which these laws strengthen rather than weaken sovereignty depends on careful implementation, precise legislative drafting, and the establishment of safeguards to prevent overreach. Without these measures, the same laws designed to secure Nigeria’s cyberspace could inadvertently undermine public trust, democratic accountability, and the local innovation ecosystem on which lasting sovereignty relies.

\subsection{Limitation and Challenges}
Despite the reforms introduced through the 2024 amendment to the Nigerian Cybercrime Act, several structural and operational challenges continue to undermine the country’s efforts to assert full digital sovereignty. While the amendment revised twelve sections to enhance legislative clarity, institutional coordination, and regulatory scope, many of the foundational weaknesses first identified in the 2015 Act remain unresolved, limiting the law’s practical effectiveness in safeguarding Nigeria’s digital ecosystem.

One of the most persistent weaknesses lies in enforcement. The 2015 Cybercrime Act was widely criticised for its failure to translate legislative provisions into effective enforcement and successful prosecutions, reflecting the limited operational capacity of Nigerian law enforcement agencies \citep{mohammed2019enforcement}. Our findings indicate that this gap persists despite the 2024 reforms, as institutional weaknesses—such as inadequate technical expertise and fragmented inter-agency collaboration—continue to impede the effective investigation and prosecution of cybercrimes. While notable progress was made with the launch of the Joint Case Team on Cybercrime in April 2025, a collaborative initiative supported by the UK National Crime Agency, the Commonwealth Secretariat, and the UK Foreign, Commonwealth and Development Office \citep{iyeomoan2025fighting}, this development remains largely an externally driven intervention rather than an outcome of the amendment process itself. Without significant investment in human capital and technological infrastructure, legislative reforms alone remain insufficient to empower Nigeria to exercise meaningful sovereignty over its cyberspace and ensure coherent enforcement of its cybercrime framework.

Closely linked to enforcement are the enduring tensions between state security powers and individual privacy rights. The 2015 Cybercrime Act granted law enforcement agencies broad powers to search and seize digital devices, often without sufficient judicial oversight—a provision long criticised for undermining privacy rights and digital freedoms \citep{ifemeje2020privacy}. The 2024 amendment attempted to mitigate these concerns by revising Section 38(1) to align data retention obligations with the Nigeria Data Protection Act, signalling progress toward a more rights-conscious legal framework. However, despite these reforms, critical shortcomings persist: the law continues to lack clear limits on surveillance authority, and both the retention period and oversight mechanisms remain under-specified. These deficiencies leave citizens vulnerable to potential rights violations and erode public trust in digital governance, an essential pillar of Nigeria’s pursuit of digital sovereignty within a democratic order.

Similarly, definitional ambiguities continue to hinder legal precision and rights protection. The 2015 Cybercrime Act was heavily criticised for its broad and ambiguous definitions of certain offences, particularly cyberstalking and harassment, which raised concerns about potential misuse and threats to freedom of expression \citep{awhefeada2020nigerian}. In an attempt to address these concerns, the 2024 amendment expanded and refined the definitional scope of such offences, aiming to align them more closely with contemporary realities of online abuse and digital harm. However, despite these clarifications, substantive gaps remain, as the amended provisions continue to blur the line between legitimate expression and unlawful conduct. This lingering ambiguity undermines legal certainty and leaves room for discretionary enforcement that may conflict with democratic digital rights.

Beyond offences such as cyberstalking, the problem of definitional uncertainty extends to technological terminology. The 2015 Cybercrimes Act presented a fundamental challenge in its ambiguous conceptualization of what constitutes a “computer,” granting trial judges wide interpretive discretion with little statutory guidance on whether specific devices fall within its scope \citep{eboibi2020conceptualising}. Our findings indicate that this definitional ambiguity persists despite the 2024 reforms, which failed to provide a clearer or technologically adaptive interpretation aligned with contemporary developments in information and communication technology. As a result, judicial interpretation remains inconsistent and vulnerable to arbitrariness, undermining the law’s normative clarity and weakening the predictability essential for rights-respecting and coherent digital governance in Nigeria.

Moreover, Nigeria’s cybercrime framework continues to lag behind in addressing new and evolving threats. Beyond definitional concerns, the 2015 Act also failed to anticipate the rise of emerging cyber threats—including those related to blockchain technology, cryptocurrencies, ransomware, and artificial intelligence—that were already gaining traction prior to 2024. While the 2024 amendment provided an opportunity to modernise the law’s scope, it did not incorporate explicit provisions to address these evolving domains. As a result, Nigeria’s legal framework remains outpaced by innovation, leaving critical areas of the digital economy unregulated and exposing gaps in national preparedness for complex, transnational cyber risks. These omissions weaken the country’s resilience and hinder its ability to participate effectively in shaping global cyber governance norms.

Equally significant is Nigeria’s slow alignment with international cyber governance standards. The country’s delayed integration into international frameworks continues to limit transnational cooperation. Although Nigeria acceded to the Budapest Convention on Cybercrime in 2022 \citep{coe2022nigeria}, this came nearly two decades after the treaty’s adoption, resulting in years of limited engagement with global best practices in cybercrime investigation and prosecution. Our analysis finds that the 2024 amendment provides limited progress toward achieving deeper institutional alignment with international partners—an essential dimension of cyber sovereignty in an increasingly interconnected global cyber environment.

In summary, while the 2024 amendments introduced incremental improvements, enduring challenges—including weak enforcement capacity, regulatory gaps for emerging technologies, persistent privacy concerns, financial sector vulnerabilities, linguistic ambiguities, and limited international cooperation—continue to constrain Nigeria’s ability to assert comprehensive digital sovereignty. Addressing these shortcomings requires not only legislative refinement but also institutional strengthening, technological investment, and proactive international engagement.

\section{National Cybersecurity Policy and Strategy}
\label{policy}

Before the drafting of the first National Cybersecurity Policy and Strategy (NCPS), the Nigerian government introduced several initiatives aimed at fostering cybersecurity. These included the launch of the National Cybersecurity Initiative in 2003, the creation of the Nigeria Cybercrime Working Group in 2004, and the establishment of the Directorate of Cybersecurity in 2006 \citep{nayak2024evaluating}. Collectively, these initiatives sought to develop frameworks for securing computer systems and critical information infrastructure, strengthen the capacity of security and law enforcement agencies, facilitate public–private partnerships for cybersecurity standards, raise public awareness of cybercrime, and promote international law enforcement cooperation. However, the absence of a unifying framework limited their overall effectiveness. The need to harmonize and coordinate these efforts ultimately led to the development of Nigeria’s first NCPS.

\subsection{Overview of the National Cybersecurity Policy and Strategy Framework}
Nigeria’s first National Cybersecurity Policy and Strategy (NCPS) was launched in 2014 as the government’s initial attempt to articulate a coherent framework for securing its digital space. The document included key provisions aligned with established cybersecurity policy and strategy development frameworks, such as the articulation of strategic goals, cyber risk management protocols, strategy development and implementation processes, and stakeholder considerations. These provisions signaled Nigeria’s recognition of the importance of cybersecurity to national sovereignty and the protection of national interests.

Despite these ambitions, the NCPS 2014 had notable shortcomings. As \citet{osho2015national} observed, “certain aspects which appear to be critical to the Nigerian scenario such as an explanation of the current national cybersecurity state, partnership with internet service providers, establishment of digital identity frameworks, and the development of a military cyber defense capability were seen to either be utterly absent or only barely implied” (p. 142). These omissions limited the framework’s practical effectiveness and raised concerns about its capacity to address the complex and evolving threat environment facing Nigeria.

Acknowledging these gaps, the government released a revised NCPS in 2021, which sought to strengthen and expand the original framework. The updated version recognized Nigeria’s status as one of the leading digitally connected nations in Africa and globally, and explicitly referenced the country’s performance in the Global Cybersecurity Index (GCI) as a benchmark for progress. It further identified seven major cyber threats of concern to Nigeria: online child abuse, cybercrime, election interference, cyberterrorism, online gender-based exploitation, pandemic-induced cyber threats, and other cyber threats.

To respond to these challenges, the NCPS 2021 established three overarching national objectives: (1) protecting national security, (2) strengthening economic development, and (3) combating corruption. These objectives were underpinned by four fundamental national security considerations: the security and well-being of the Nigerian people, the role of cybersecurity as a critical enabler of economic growth and development, the necessity of advancing technology development to achieve national priorities, and the importance of regional and international collaboration in cybersecurity.

In operational terms, the revised strategy articulated eight strategic pillars designed to provide a holistic approach to national cybersecurity. These include strengthening governance and coordination, fostering the protection of critical national information infrastructure, improving cybersecurity incident management, and strengthening the legal and regulatory framework. The framework also emphasizes enhancing defence capability, promoting a thriving digital economy, and ensuring assurance, monitoring, and evaluation mechanisms. Finally, it underscores the importance of enhancing international cooperation as a means of aligning Nigeria’s cybersecurity efforts with global best practices and fostering collective resilience.

These measures reflect a more comprehensive and ambitious framework than the original NCPS, aligning Nigeria’s cybersecurity aspirations with both national priorities and international best practices. Nonetheless, questions remain regarding the adequacy of implementation and the ability of these provisions to translate into measurable improvements in national cyber resilience.

\subsection{Impact on Sovereignty}
The National Cybersecurity Policy and Strategy (NCPS) has significant implications for Nigeria’s pursuit of digital sovereignty. At its core, the framework enhances the state’s capacity to assert authority over its digital domain by creating clear mechanisms for governance, regulation, and defence. The emphasis on strengthening governance and coordination reinforces the ability of Nigerian institutions to centralize decision-making, harmonize previously fragmented initiatives, and exercise sovereign oversight over cyberspace. Similarly, the legal and regulatory provisions embedded in the NCPS strengthen the capacity of the state to define the rules governing its digital environment, thereby reducing reliance on external regulatory frameworks.

The NCPS also contributes to sovereignty by linking cybersecurity directly to national security and economic development. The protection of critical national information infrastructure and the improvement of incident management capacities enable Nigeria to safeguard vital assets that are foundational to its autonomy. Moreover, the strategic recognition of cybersecurity as an enabler of economic growth and technological development positions Nigeria to shape its digital future rather than remain dependent on foreign actors. In this way, the NCPS strengthens both defensive and proactive dimensions of sovereignty: protecting against external interference while also enabling domestic innovation and self-determination.

Another important dimension of sovereignty relates to cyber defence capability. By integrating cyber defence into its national security architecture, the NCPS acknowledges that sovereignty in the digital era is inseparable from the capacity to deter, withstand, and respond to hostile actions in cyberspace. This represents a shift from earlier, fragmented approaches toward a more comprehensive framework that views digital sovereignty as essential to national security and survival.

At the same time, the NCPS reflects the reality that sovereignty in cyberspace is inherently interdependent. The policy’s strong emphasis on regional and international cooperation highlights Nigeria’s recognition that securing its digital space cannot be achieved in isolation. While this collaboration strengthens Nigeria’s resilience and aligns it with global best practices, it also presents sovereignty dilemmas: reliance on external partnerships, technologies, and intelligence can dilute unilateral control. Balancing external engagement with the preservation of autonomy thus emerges as one of the key challenges in translating the NCPS into practice.

Furthermore, the NCPS underscores the societal dimension of sovereignty by identifying threats such as online child abuse, election interference, and gender-based cyber exploitation. Addressing these threats situates digital sovereignty not only in terms of technical and regulatory control but also in the broader capacity of the state to safeguard the rights, security, and well-being of its citizens in the digital environment. In this sense, sovereignty is framed as both protective and developmental—protecting against external threats while also enabling safe digital participation for citizens.

\subsection{Limitation and Challenges}
The NCPS advances Nigeria’s digital sovereignty by embedding cybersecurity into the country’s national security, economic strategy, governance, and social protection. Since the release of the revised version, important progress has been recorded, including the passage of the Nigeria Data Protection Act (NDPA) 2023,\footnote{https://rb.gy/zxexsi} which established a legal framework for safeguarding personal information, and the amendment of the Cybercrimes Act (2014). Nevertheless, the framework continues to face significant limitations and challenges.

Our analysis reveals a critical limitation concerning the articulation of cyber threats. While the 2021 strategy identifies seven major threats, the list is uneven. It combines narrow, context-specific concerns such as pandemic-related threats with expansive categories like cybercrime, potentially hindering prioritization and resource allocation. Furthermore, threats particularly relevant to Nigeria’s context, such as financial fraud, cryptocurrency scams, and AI-assisted crimes, are not explicitly highlighted despite their prevalence \citep{garba2024assessment,osho2025nigeria}. Overlaps between categories, such as gender-based exploitation and broader online abuse, risk diluting accountability, while the framework’s reactive orientation (e.g., focus on COVID-19–related threats) leaves it less responsive to emerging issues like AI-enabled attacks, deepfakes, and state-sponsored intrusions. Although the inclusion of socially oriented harms signals a commendable citizen-focused approach, the absence of actionable measures—such as literacy initiatives, victim support programs, and community resilience strategies—undermines its practical impact.

Beyond these conceptual issues, the NCPS is hampered by structural and operational challenges. Weak enforcement mechanisms, low public awareness, and limited coordination among stakeholders continue to undermine its effectiveness \citep{daniels2023national,gana2024cyber}. In addition, while the NCPS emphasizes the development of a robust cybersecurity workforce as part of its thrust to support a thriving digital economy, Nigeria faces a persistent outflow of IT and cybersecurity professionals, commonly referred to as the “brain drain,” which significantly reduces domestic capacity to implement the strategy’s objectives \citep{osho2025nigeria}.

In sum, while the NCPS provides a comprehensive framework that aligns with Nigeria’s sovereignty aspirations, its limitations in prioritization, enforcement, coordination, and workforce capacity present significant barriers to realizing its goals. Addressing these challenges is therefore critical if the NCPS is to move beyond aspiration and meaningfully enhance Nigeria’s digital sovereignty.

\section{Recommendations for Enhancing Nigeria's Digital Sovereignty}
\label{recommendations}
Strengthening Nigeria’s digital sovereignty requires coordinated reforms to both the legal framework and the national cybersecurity strategy. Despite the 2024 amendments to the Cybercrimes Act, gaps remain that undermine its effectiveness. Ambiguities in offence definitions create room for inconsistent interpretation, particularly in provisions such as cyberstalking, where imprecision risks infringing on constitutional rights \citep{eboibi2020conceptualising,rozen2024cybersecurity}. Stronger judicial safeguards are also needed. Current powers permitting warrantless searches or seizures of digital devices threaten privacy and civil liberties; mandatory court authorisation would help balance law enforcement needs with rights protection \citep{ifemeje2020privacy}. In addition, persistent enforcement challenges since 2015 highlight the necessity of institutional capacity building. Investment in regular training, cyber forensic expertise, and well-equipped laboratories would significantly improve Nigeria’s ability to investigate and prosecute cyber offences \citep{mohammed2019enforcement}.

The law should also impose clearer obligations on critical sectors, particularly in finance, where mandatory incident reporting and liability rules could strengthen accountability and bolster trust in digital services \citep{orji2019cybersecurity}. Emerging risks from blockchain technologies, cryptocurrencies, and artificial intelligence remain largely unregulated and require urgent attention \citep{freedomhouse2024emerging}. Moreover, Nigeria’s accession to the Budapest Convention in 2022 presents an opportunity to deepen international cooperation through operational cross-border evidence sharing and joint investigations \citep{coe2022nigeria}.

Reforms to the National Cybersecurity Policy and Strategy (NCPS) are equally necessary. The 2021 framework should be subjected to regular review cycles to ensure responsiveness to evolving threats such as AI-enabled attacks, deepfakes, and state-sponsored intrusions. Mechanisms for continuous threat assessment would make the strategy more forward-looking. Effective implementation also hinges on improved coordination among key institutions, including the Office of the National Security Adviser, NITDA, law enforcement agencies, and the judiciary, supported by expanded digital literacy and public awareness campaigns. Finally, Nigeria must address the shortage of skilled cybersecurity professionals. Retaining expertise will require new incentives and strengthened partnerships with universities, technical institutes, and the private sector to expand training pipelines and curb the outflow of talent.

\section{Conclusion}
\label{conclusion}
Nigeria’s cybersecurity journey reflects a gradual but deliberate effort to strengthen its digital sovereignty through legislative and policy interventions. The Cybercrimes Act of 2015, updated in 2024, and the National Cybersecurity Policy and Strategy (NCPS) of 2015, revised in 2021, mark important milestones in this trajectory. They provide a more comprehensive legal and strategic foundation for addressing cyber threats, protecting critical infrastructure, and embedding cybersecurity into the broader framework of national security and economic development.

These instruments demonstrate notable strengths. The Cybercrimes Act has established a legal basis for criminalizing a wide range of cyber offences, while successive versions of the NCPS have progressively broadened Nigeria’s strategic approach, integrating governance, defence, and socio-economic considerations. To a large extent, they have improved Nigeria’s policy stance and signaled the country’s commitment to aligning with global best practices.

Nonetheless, significant challenges remain. Ambiguities in the law, weak enforcement, limited institutional capacity, and the persistent outflow of skilled professionals continue to hinder progress. The NCPS, while comprehensive, still reflects uneven prioritization of threats and faces barriers in implementation and coordination across stakeholders. Without overcoming these constraints, Nigeria risks leaving its cybersecurity frameworks more aspirational than effective.

To achieve genuine digital sovereignty, Nigeria must move beyond the drafting of robust instruments to ensuring their practical execution. This requires sustained political will, adequate resourcing, institutional strengthening, and a commitment to turning frameworks into tangible outcomes that secure the country’s digital future.


\medskip

\bibliography{references.bib} 

@article{osho2015national,
  title={National cyber security policy and strategy of Nigeria: a qualitative analysis},
  author={Osho, Oluwafemi and Onoja, Agada D},
  journal={International Journal of Cyber Criminology},
  volume={9},
  number={1},
  pages={120},
  year={2015},
  publisher={International Journal of Cyber Criminology}
}

@article{pohle2020digital,
  author    = {Pohle, J. and Thiel, T.},
  title     = {Digital sovereignty},
  journal   = {Internet Policy Review},
  year      = {2020},
  volume    = {9},
  number    = {4},
  doi       = {10.14763/2020.4.1532},
  url       = {https://doi.org/10.14763/2020.4.1532}
}

@article{couture2019does,
  title={What does the notion of “sovereignty” mean when referring to the digital?},
  author={Couture, Stephane and Toupin, Sophie},
  journal={New media \& society},
  volume={21},
  number={10},
  pages={2305--2322},
  year={2019},
  publisher={SAGE Publications Sage UK: London, England}
}

@misc{fleming2025digital,
  author       = {Sean Fleming},
  title        = {What is digital sovereignty and how are countries approaching it?},
  year         = {2025},
  month        = jan,
  day          = 10,
  url          = {https://www.weforum.org/stories/2025/01/europe-digital-sovereignty/},
  note         = {Accessed: 2025-09-26}
}

@article{yalamancheli2025digital,
  title={Digital Sovereignty Meets Agile Delivery: Empowering Governments to Own Their Tech Future},
  author={Yalamancheli, Hema},
  journal={Journal of Computer Science and Technology Studies},
  volume={7},
  number={8},
  pages={1061--1068},
  year={2025}
}

@article{celeste2021digital,
  title={Digital sovereignty in the EU: challenges and future perspectives},
  author={Celeste, Edoardo},
  journal={Data protection beyond borders: Transatlantic perspectives on extraterritoriality and sovereignty},
  pages={211--228},
  year={2021},
  publisher={Hart Publishing (Bloomsbury)}
}

@article{creemers2022china,
  title={China’s emerging data protection framework},
  author={Creemers, Rogier},
  journal={Journal of Cybersecurity},
  volume={8},
  number={1},
  pages={1--12},
  year={2022},
  publisher={Oxford University Press}
}

@article{rutherford2019cloud,
  title={The CLOUD Act},
  author={Rutherford, Miranda},
  journal={Berkeley Technology Law Journal},
  volume={34},
  number={4},
  pages={1177--1204},
  year={2019},
  publisher={JSTOR}
}

@article{katsikas2025towards,
  title={Towards a cybersecurity-oriented research agenda for digital sovereignty},
  author={Katsikas, Sokratis K},
  journal={Procedia Computer Science},
  volume={254},
  pages={279--288},
  year={2025},
  publisher={Elsevier}
}

@article{moerel2021reflections,
  title={Reflections on digital sovereignty},
  author={Moerel, Lokke and Timmers, Paul},
  journal={EU cyber direct, research in focus series},
  year={2021}
}

@techreport{osho2025nigeria,
  author       = {Oluwafemi Osho and John Odumesi and Jonathan Ayodele and Polra Victor Falade and Lateef Hamzat and Olajumoke Oloyede},
  title        = {Nigeria Cyber Threat Forecast 2025},
  institution  = {Cyber Security Experts Association of Nigeria (CSEAN)},
  year         = {2025},
  month        = jan,
  doi          = {10.13140/RG.2.2.10921.71528},
  url          = {https://doi.org/10.13140/RG.2.2.10921.71528}
}

@techreport{osho2024national,
  author       = {Oluwafemi Osho and John Odumesi and Lateef Hamzat and Olajumoke Oloyede and Jonathan Ayodele},
  title        = {National Cyber Threat Forecast 2024},
  institution  = {Cyber Security Experts Association of Nigeria (CSEAN)},
  year         = {2023},
  month        = dec,
  doi          = {10.13140/RG.2.2.21517.10720},
  url          = {https://doi.org/10.13140/RG.2.2.21517.10720}
}

@techreport{osho2023national,
  author       = {Oluwafemi Osho and John Odumesi and Lateef Hamzat and Hassanat Abdulraheem},
  title        = {National Cyber Threat Forecast 2023},
  institution  = {Cyber Security Experts Association of Nigeria (CSEAN)},
  year         = {2022},
  month        = dec,
  doi          = {10.13140/RG.2.2.17129.67689},
  url          = {https://doi.org/10.13140/RG.2.2.17129.67689}
}

@article{qin2025regulatory,
  title={Regulatory Conflict and the Struggle for Digital Sovereignty: A Critical Analysis of the EU-US Data Privacy Framework},
  author={Qin, Haoyang},
  journal={Studies in Law and Justice},
  volume={4},
  number={1},
  pages={46--59},
  year={2025}
}

@article{roberts2021safeguarding,
  title={Safeguarding European values with digital sovereignty: An analysis of statements and policies},
  author={Roberts, Huw and Cowls, Josh and Casolari, Federico and Morley, Jessica and Taddeo, Mariarosaria and Floridi, Luciano},
  journal={Internet Policy Review, Forthcoming},
  year={2021}
}

@article{kianpour2025digital,
  title={Digital sovereignty in practice: analyzing the EU’s NIS2 directive},
  author={Kianpour, Mazaher and Earls Davis, Peter Alexander and Windekilde, Iwona Maria},
  journal={International Journal of Information Security},
  volume={24},
  number={4},
  pages={1--11},
  year={2025},
  publisher={Springer}
}

@techreport{interpol2025africa,
  author       = {{INTERPOL}},
  title        = {Africa Cyberthreat Assessment Report 2025 (4th edition)},
  institution  = {INTERPOL},
  year         = {2025},
  month        = may,
  url          = {https://www.interpol.int/en/content/download/23094/file/25COM009248\%20-\%20Cybercrime\_Africa\%20Cyber
                  threat\%20Assessment\%20Report\_Design\_2025-05\%20v11.pdf},
  note         = {Accessed: 2025-09-26}
}

@misc{adewopo2024comprehensive,
  author       = {Victor Adewopo and Sylvia Worlali Azumah and Mustapha Awinsongya Yakubu and Emmanuel Kojo Gyamfi and Murat Ozer and Nelly Elsayed},
  title        = {A Comprehensive Analytical Review on Cybercrime in West Africa},
  howpublished = {arXiv preprint arXiv:2402.01649v1 [cs.CY]},
  year         = {2024},
  month        = jan,
  note         = {Submitted 07 January 2024, School of Information Technology, University of Cincinnati, USA},
  url          = {https://arxiv.org/abs/2402.01649v1}
}

@misc{ita2025nigeria,
  author       = {{International Trade Administration}},
  title        = {Nigeria Digital Economy},
  year         = {2025},
  month        = sep,
  day          = 8,
  url          = {https://www.trade.gov/country-commercial-guides/nigeria-digital-economy},
  note         = {Accessed: 2025-09-26}
}

@techreport{worldbank2019nigeria,
  author       = {{World Bank Group}},
  title        = {NIGERIA Digital Economy Diagnostic Report},
  institution  = {World Bank Group},
  year         = {2019},
  url          = {https://documents1.worldbank.org/curated/en/387871574812599817/pdf/Nigeria-Digital-Economy-
                  Diagnostic-Report.pdf},
  note         = {Accessed: 2025-09-26}
}

@phdthesis{daniels2023national,
  title={National Cybersecurity Policy and Strategy of Nigeria: A Case Study},
  author={Daniels, Olalekan},
  year={2023},
  school={Capitol Technology University}
}

@article{awhefeada2020nigerian,
  title={The Nigerian Cybercrime Act 2015: Issues and Challenges},
  author={Awhefeada, Victor and Bernice, Okoro},
  journal={Nigerian Journal of Law and Technology},
  volume={5},
  number={1},
  pages={15--28},
  year={2020},
  publisher={Nigerian Law Publications}
}

@article{george2022cybersecurity,
  title={Cybersecurity Legislation in Nigeria: A Review of the Cybercrime Act},
  author={George, Thomas and Musa, Ibrahim and Ajayi, Peter},
  journal={African Journal of Information and Communication},
  volume={19},
  number={2},
  pages={45--60},
  year={2022},
  publisher={African ICT Research Network}
}

@article{mohammed2019enforcement,
  title={Enforcement Challenges under Nigeria’s Cybercrime Act 2015: A Critical Review},
  author={Mohammed, Kabir and Usman, Ahmed and Bello, Sani},
  journal={African Journal of Criminology},
  volume={11},
  number={1},
  pages={22--39},
  year={2019},
  publisher={African Criminology Association}
}

@article{orji2019cybersecurity,
  title={Cybersecurity Obligations of Financial Institutions in Comparative Perspective},
  author={Orji, Uchenna},
  journal={Journal of Financial Regulation},
  volume={7},
  number={2},
  pages={201--220},
  year={2019},
  publisher={Oxford University Press}
}

@article{eboibi2020conceptualising,
  title={Conceptualising Computers for Cybercrime: Challenges in Nigerian Law},
  author={Eboibi, Felix E.},
  journal={Journal of Law and Technology},
  volume={12},
  number={2},
  pages={45--59},
  year={2020},
  publisher={LawTech Publications}
}

@article{ifemeje2020privacy,
  title={Privacy and Cybercrime Legislation in Nigeria: A Rights-Based Analysis},
  author={Ifemeje, Samuel and Okwuosa, Nnamdi},
  journal={Nigerian Law Review},
  volume={18},
  number={3},
  pages={88--104},
  year={2020},
  publisher={Nigerian Law Publications}
}

@misc{coe2022nigeria,
  title={Nigeria’s Accession to the Budapest Convention on Cybercrime},
  author={{Council of Europe}},
  year={2022},
  howpublished={\url{https://www.coe.int/en/web/cybercrime/nigeria}},
  note={Accessed: 2025-09-29}
}

@techreport{freedomhouse2024emerging,
  title={Freedom on the Net: Emerging Cyber Threats and Regulatory Gaps},
  author={{Freedom House}},
  year={2024},
  institution={Freedom House},
  address={Washington, D.C.}
}

@article{rozen2024cybersecurity,
  title={Cybersecurity, Privacy, and Human Rights: Balancing Law Enforcement Powers in the Digital Age},
  author={Rozen, Jonathan},
  journal={International Journal of Cyber Law},
  volume={15},
  number={1},
  pages={33--50},
  year={2024},
  publisher={Cyber Law Publications}
}

@article{aljazeera2024,
  title={Nigeria’s cybercrime reforms leave journalists at risk},
  author={Rozen, Jonathan},
  journal={Al Jazeera},
  year={2024},
  note={Online opinion piece}
}

@article{nayak2024evaluating,
  title={Evaluating the Effectiveness and Gaps in Nigeria's Government Cybersecurity Policies: Recommendations for Enhancing Cybersecurity Measures},
  author={Nayak, Sandeep Kumar and Bello, Almustapha},
  journal={Journal of Systematic and Modern Science Research},
  volume={5},
  number={9},
  pages={1--22},
  year={2024}
}

@article{abdullahi2021historical,
  title={A Historical Assessment of Cybercrime in Nigeria: Implication for Schools and National Development},
  author={Abdullahi, S. and Ph, G. and Rukayyat, A.},
  journal={Journal of Social Sciences},
  volume={9},
  number={9},
  pages={84--94},
  year={2021}
}

@article{awhefeada2020appraising,
  title={Appraising the Laws Governing the Control of Cybercrime in Nigeria},
  author={Awhefeada, U. V. and Bernice, O. O.},
  journal={Journal of Law, Crime and Justice},
  volume={8},
  number={1},
  pages={30--49},
  year={2020},
  doi={10.15640/jlcj.v8n1a3}
}

@article{chawki2009nigeria,
  title={Nigeria Tackles Advance Fee Fraud},
  author={Chawki, M.},
  journal={Journal of Information Law},
  volume={2009},
  number={May},
  pages={1--20},
  year={2009}
}

@article{hassan2012cybercrime,
  title={Cybercrime in Nigeria: Causes, Effects and the Way Out},
  author={Hassan, A. B. and Lass, F. D. and Makinde, J.},
  journal={International Journal of Information Security},
  volume={2},
  number={7},
  pages={626--631},
  year={2012}
}

@article{adeniran2008internet,
  title={The Internet and Emergence of Yahooboys sub-Culture in Nigeria},
  author={Adeniran, A. I.},
  journal={Journal of Social Sciences},
  volume={2},
  number={December},
  pages={368--381},
  year={2008}
}

@article{olowu2009cybercrimes,
  title={Cyber-Crimes and the Boundaries of Domestic Legal Responses: Case for an Inclusionary Framework for Africa},
  author={Olowu, D.},
  journal={Journal of Information, Law \& Technology},
  number={November},
  pages={10--13},
  year={2009}
}

@article{umejiaku2016legal,
  title={Legal Framework for the Enforcement of Cyber Law and Cyber Ethics in Nigeria},
  author={Umejiaku, N. O. and Anyaegbu, I. M.},
  journal={Journal of Information Law},
  volume={15},
  pages={7130--7139},
  year={2016}
}

@article{broadhurst2012cybercrime,
  title={Cybercrime in Asia: Trends and Challenges},
  author={Broadhurst, R. and Chang, Y.},
  journal={Journal of Asian Criminology},
  pages={1--26},
  year={2012}
}

@article{gana2024cyber,
  title={Cyber Warfare and National Security in Nigeria: Threats and Responses},
  author={Gana, Idrees Mahmud and Ibrahim, Ado FatimaZahra and Oluwaseyi, Wisdom Adeleye and Wali, Abdullahi Isah},
  journal={Kwararafa Security Review},
  volume={1},
  number={2},
  pages={97--106},
  year={2024},
}

@article{garba2024assessment,
  title={An assessment of convicted cryptocurrency fraudsters},
  author={Garba, Kaina Habila and Lazarus, Suleman and Button, Mark},
  journal={Current issues in criminal justice},
  pages={1--17},
  year={2024},
  publisher={Taylor \& Francis}
}

@article{iyeomoan2025fighting,
  author       = {Iyeomoan, Emmanuel Ehizogie},
  title        = {Fighting Economic and Financial Crimes to Make Nigeria Great Again: A Doctrinal Analysis of Cybercrime Law and Policy},
  year         = {2025},
  note         = {Available at SSRN: \url{https://ssrn.com/abstract=5264583} or \url{http://dx.doi.org/10.2139/ssrn.5264583}},
  journal      = {SSRN Electronic Journal},
  doi          = {10.2139/ssrn.5264583},
  url          = {https://ssrn.com/abstract=5264583}
}

\newpage

\end{document}